\documentclass[10pt,conference,letterpaper]{IEEEtran}

\usepackage[
  letterpaper,
  textwidth=178mm,
  textheight=229mm,
  left=19mm,
  top=25mm,
  columnsep=6mm,
  headheight=0pt,
  headsep=0pt
]{geometry}

\usepackage{cite}
\usepackage{amsmath,amssymb,amsfonts}
\usepackage{graphicx}
\usepackage{url}
\usepackage{booktabs}
\usepackage[hidelinks]{hyperref}
\graphicspath{{fig/}}
\usepackage{tikz}
\usetikzlibrary{arrows.meta, positioning, calc}

\begin{document}

\title{A SIGNAL-PROCESSING CASCADE LINKING SATELLITE HEAT EXPOSURE, GENOMIC VARIATION, AND MORTALITY RISK}

\author{
\IEEEauthorblockN{Usman Nazir, Sara Khalid}
\IEEEauthorblockA{PHI Lab, NDORMS\\
University of Oxford, Oxford, United Kingdom\\
Email: \{usman.nazir, sara.khalid\}@ndorms.ox.ac.uk}
}

\maketitle

\begin{abstract}
Heat-related mortality varies among individuals under similar environmental conditions, yet environmental exposure and genetic susceptibility are typically modelled separately. We propose a signal-processing cascade linking satellite-derived land-surface temperature to sequence-dependent heat-shock response and mortality risk. The framework combines distributed-lag filtering of thermal exposure, nonlinear HSF1 activation, cross-correlation with an HSF1 position-weight-matrix kernel, and Hill-model mapping of sequence-dependent binding changes to regulatory response. Applied to 52 heat-response genes and 157 naturally occurring variants, the framework identifies 21 heat-shock-element arrays across 17 genes. Variants in canonical arrays are significantly enriched for predicted binding disruption ($p=3\times10^{-4}$). Propagating these effects through the cascade yields a peak predicted mortality-risk multiplier of $1.68\times$ at $40$--$42\,^{\circ}\mathrm{C}$. The framework demonstrates how environmental and genomic signals can be integrated to model inter-individual variation in heat-related risk.
\end{abstract}


\section{Introduction}

Extreme heat is a major and growing public-health hazard~\cite{WHOHeatHealth}. Heat stress can directly cause heat
exhaustion and heat stroke and can also exacerbate cardiovascular,
renal, respiratory, and other pre-existing
conditions~\cite{WHOHeatHealth}. Large heatwave events have
consequently produced substantial excess mortality, including more
than 70,000 deaths during the 2003 European heatwave and an
estimated 61,672 heat-related excess deaths in Europe during the
summer of 2022~\cite{WHOHeatHealth}. Climate change is increasing
population exposure to extreme heat and the associated burden of heat-related mortality~\cite{vicedoCabrera2021}.

The health effects of heat, however, are not uniform across individuals. Age, underlying disease, hydration, occupation, and
other physiological and environmental factors modify vulnerability to heat~\cite{WHOHeatHealth}. At the cellular level, an important
protective mechanism is the heat-shock response, coordinated by heat
shock factor 1 (HSF1). Under proteotoxic stress, HSF1 trimerizes,
binds heat-shock elements (HSEs) in the promoters of target genes,
and activates the expression of molecular chaperones that help
maintain proteostasis~\cite{neudegger2016structure, yamamoto2009differential}.
% When the outside temperature is very high, the body cannot dump heat
% fast enough. Core temperature climbs above the safe range of
% $37\,^\circ$C and, above roughly $41\,^\circ$C, the machinery inside
% the cell starts to break. Kidneys, heart, brain and blood all
% suffer. This is why hospitals in a heatwave see spikes in kidney
% failure, stroke and sudden cardiac arrest.

% Two people standing in the same $42\,^\circ$C sun for the same time
% do not have the same risk of dying. Age, hydration and cardiovascular
% fitness matter, but so does a hidden variable inherited at birth.
% Every cell contains an emergency protection system that switches on
% when the cell heats up, and a cell heats up only when the body's
% core temperature climbs, and the body's core temperature climbs only
% when the outside weather is hot enough for long enough. So
% \emph{hot weather $\to$ hot body $\to$ hot cell $\to$ switch on}. The
% switch itself is a short piece of DNA. Most people carry a fully
% functional version. A small minority carry a version with a
% one-letter typo that weakens the switch. Under normal weather the
% typo does not matter; during a heatwave it matters a lot.
\subsection{Relevant exposure time scales}

Heat-related health effects occur over multiple time scales.
Acute heat exposure can raise core temperature and trigger heat illness within hours, while repeated hot days can impose cumulative
cardiovascular and renal strain. Epidemiological studies also show
that temperature-related mortality can be delayed beyond the day of exposure, motivating the use of distributed-lag models \cite{vicedoCabrera2021}. We therefore represent recent thermal exposure using a finite distributed-lag window rather than treating each day's temperature independently.

\begin{figure*}
\centering
\includegraphics[scale=0.22]{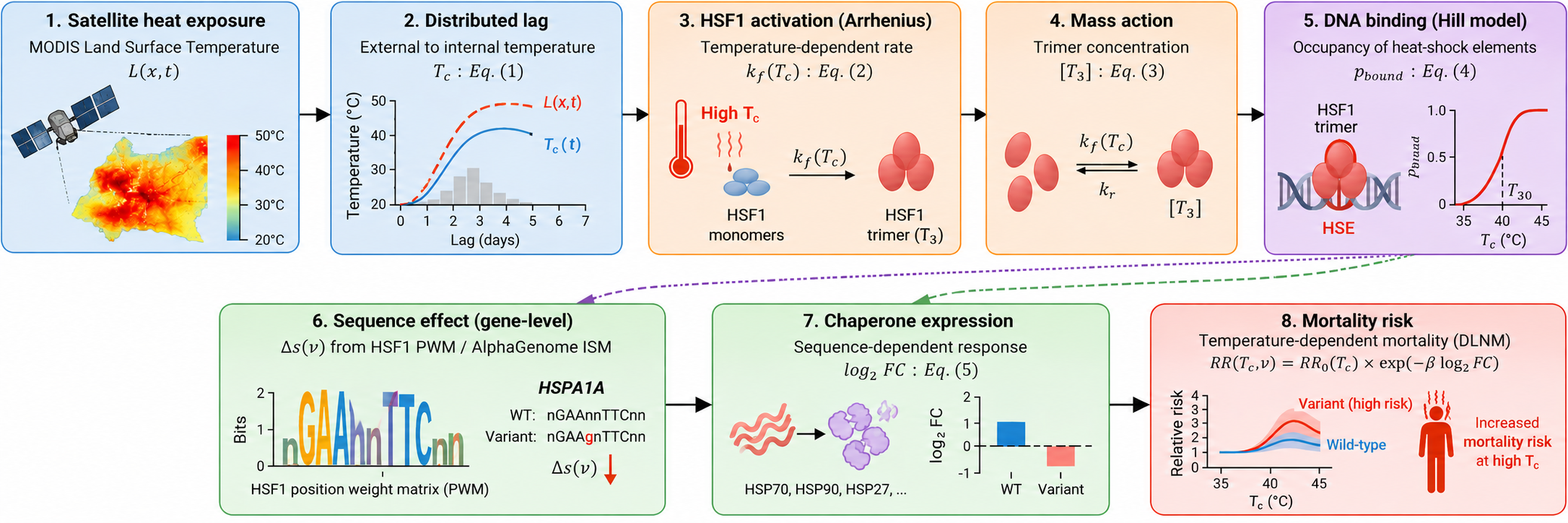}
\caption{The four-layer model. Blue: satellite and epidemiology.
Orange: biophysics. Purple: binding. Green: sequence and expression.
Red: outcome. Dashed and dotted arrows show that the sequence-level
$\Delta s(v)$ propagates through $K_d$ into occupancy and thence
into $\log_2$FC and mortality.}
\label{fig:framework}
\end{figure*}

\subsection{Contributions}
Existing risk maps from satellite data give one number per
neighbourhood (how hot it will be) and stop. They do not say
who in the neighbourhood is most at risk. We connect the satellite
measurement to the personal switch by treating the whole pathway as
a cascade of four SP operators. Nothing in it needs a solver more
sophisticated than a convolution or a pointwise transform.

\section{Method: a Four-Step Cascade}

The mapping from a MODIS pixel value to a per-individual mortality
change is written as
\begin{equation}
L(x,t)\;\xrightarrow{h_1}\;T_c\;\xrightarrow{h_2}\;[T_3]
\;\xrightarrow{h_3}\;P_{\text{bound}}\;\xrightarrow{h_4}\;\Delta\text{mortality}.
\label{eq:cascade}
\end{equation}

{what each symbol means:}
\begin{itemize}
\item $L(x,t)$: outside temperature over your $1$\,km neighbourhood
      $x$ on day $t$, measured by the MODIS satellite.
\item $T_c$: your estimated internal body temperature.
\item $[T_3]$: how much of your body's protection switch is turned
      on.
\item $P_{\text{bound}}$: how firmly the switch is grabbing the DNA
      pattern that turns on the protection genes.
\item $\Delta\text{mortality}$: how much your risk of dying changed
      because of the heat.
\end{itemize}

Figure~\ref{fig:framework} summarizes the complete pipeline. The blue blocks represent the environmental exposure and epidemiological components, the orange blocks describe heat-induced biophysical activation, and the purple and green blocks represent sequence-specific HSF1 binding and gene expression. The red block gives the final mortality output. A genetic variant changes the sequence score $\Delta s(v)$, which alters the binding constant $K_d$, HSF1 occupancy, chaperone expression, and ultimately the predicted mortality risk. The dashed and dotted arrows in Fig.~\ref{fig:framework} show this sequence-dependent pathway.

\subsection{Step 1 ($h_1$): read the heat}
NASA's MODIS satellite measures the temperature of every $1\,$km
patch of the Earth every day, giving a time series $L(x,t)$. Core
body temperature responds with a delay because the body does not
warm up instantly. The delay is a finite-impulse-response filter
of length $L\!\le\!21$ days,
\begin{equation}
T_c(x,t) \;=\; 37 + \sum_{\ell=0}^{L}\!w_\ell
\bigl[L(x,t-\ell)-L_{\mathrm{ref}}(x)\bigr],
\label{eq:tcore}
\end{equation}
i.e.\ $T_c-37 = w \ast (L-L_{\mathrm{ref}})$. The taps $w_\ell$
are read off the distributed-lag non-linear model of
\cite{vicedoCabrera2021}. $L_{\mathrm{ref}}(x)$ is the pixel-specific
$95\text{th}$ percentile summer temperature over the $1981$--$2010$
baseline.

\subsection*{How long does someone need to be exposed?}
The filter captures three time-scales at once. Minutes to a few
hours: direct heat stress at $\sim\!40\,^\circ$C pushes core past
$39\,^\circ$C within $15$--$30$~minutes of exertion, and the
molecular switch begins to turn on within about $30$~minutes of the
body being pushed past normal. One to seven days: cumulative
cardio-renal strain accumulates even without a single dramatic hour,
which is why heatwave mortality is reported over multi-day windows.
Up to $21$ days: delayed deaths from organ damage, infection and
medication interactions. Our filter uses the $0$--$21$ day window
adopted directly from the epidemiology literature.

\subsection{Step 2 ($h_2$): read the switch state}
The heat-shock factor HSF1 is inactive as a monomer at
$37\,^\circ$C. Above a temperature $T_m$ it flips cooperatively to
the active trimer. The population-level fraction that is active
follows a saturating Boltzmann,
\begin{equation}
f_{T_3}(T_c) \;=\; \bigl[1+\exp\!\bigl((T_m{-}T_c)/a\bigr)\bigr]^{-1},
\label{eq:boltz}
\end{equation}
with $T_m\!=\!42\,^\circ$C and slope $a\!=\!0.7\,^\circ$C from
in-vitro measurements. This is the second nonlinearity in the
cascade.

\subsection{Step 3 ($h_3$): read the switch sequence}
The switch is a $\sim15$-bp DNA pattern in the promoters of the
body's chaperone genes. Whether a person's switch is intact reduces
to a 1-D cross-correlation between their promoter DNA and a log-odds
kernel. Encode the DNA one-hot as $X(p,b)\!\in\!\{0,1\}^{4\times N}$
and take the kernel $K(i,b) = \log_2(f_{i,b}/0.25)$ from the JASPAR
HSF1 position weight matrix (PWM; identifier MA0486.2, length $13$,
information content $18.7\,$bits). The scanner is
\begin{equation}
s(p) = \max\!\Bigl(\sum_{i,b}\!K(i,b)\,X(p{+}i,b),\;
\sum_{i,b}\!K(i,b)\,\bar X(p{-}i,b)\Bigr),
\label{eq:xcorr}
\end{equation}
where $\bar X$ is the reverse complement. A valid switch is any
alternating run of at least three pentamer hits with inter-unit gap
set $\{(0,0),(5,0),(0,5),(5,5)\}$, i.e.\ the three known
architectures documented biochemically \cite{yamamoto2009differential,ota2013evolutionarily}.
The scanner sees $1$ hit per $100\,$kb of random DNA (Poisson tail:
$p<10^{-12}$ per positive). A one-letter typo carried by a person
shifts the score by $\Delta s(v)$; this is what we call a
$\Delta$PWM. Occupancy of the switch at score $s$ is Hill,
\begin{equation}
P_{\mathrm{bound}}(s,[T_3]) =
\frac{[T_3]^{n}}{K_d(s)^{n}+[T_3]^{n}},\;
K_d(s) = K_{d,0}\,2^{-s},
\label{eq:hill}
\end{equation}
with cooperativity $n \in \{3, 2, 1.4\}$ for the perfect, gap and
step architectures.

\subsection{Step 4 ($h_4$): read the risk}
Chaperone induction $\log_2$FC is proportional to the occupancy
gap between exposed and baseline, and mortality risk is modulated
multiplicatively by
$\exp(-\beta\,[\log_2\text{FC}_{\text{ALT}}-\log_2\text{FC}_{\text{REF}}])$
with $\beta\!=\!0.35$ from an HSPA1B haplotype cohort study
\cite{alele2022next}.

\subsection{What the composed cascade actually does}
\begin{figure*}[t]
\centering
\includegraphics[width=\linewidth]{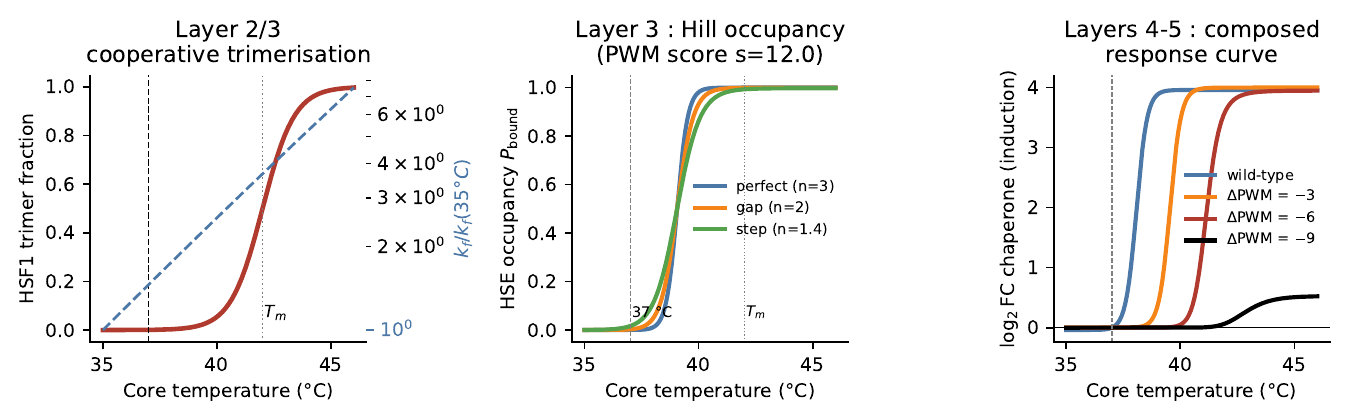}
\caption{Behaviour of the four-step cascade
(Eqs.~\ref{eq:tcore}--\ref{eq:hill}).
\textbf{Left:} Step~2, cooperative trimer fraction (red) compared
with the underlying Arrhenius rate constant $k_f$ (blue, dashed).
\textbf{Centre:} Step~3, occupancy of a canonical $s\!=\!12$ switch
under the three architectures (perfect, gap, step).
\textbf{Right:} Steps~3--4, composed temperature response of chaperone
$\log_2$FC for wild-type and for progressively disrupted variants
($\Delta$PWM $=-3,-6,-9$). A single $6$-bit loss of information at
the switch is enough to abolish the induction.}
\label{fig:math}
\end{figure*}

Figure~\ref{fig:math} shows the behaviour of the composed
Eqs.~(\ref{eq:tcore})--(\ref{eq:hill}) as a function of core body
temperature. The trimer fraction (Step~2) is essentially zero
below $40\,^\circ$C and saturates by $44\,^\circ$C. The Hill
occupancy under the three switch architectures (Step~3) diverges
sharply through the same band. The composed $\log_2$FC of chaperone
induction (Step~4) shows a clear dose-response to $\Delta$PWM: a
single $6$-bit loss of information at the switch is enough to
abolish the induction. In other words, the temperature-blind
sequence perturbation and the sequence-blind temperature step meet
at Step~3, and a defect in either one shows up in the same output.

\section{Data and Numbers}

We take a panel of $52$ heat-response genes (the union of $39$
curated HSF1Base high-confidence targets \cite{kovacs2019hsf1base} and $13$
climate-physiology genes). The composition is broken down by
category in Table~\ref{tab:panel}. Promoter DNA is pulled live
from Ensembl~112 (hg38), $\pm5$\,kb around the transcription-start
site.

Every bi-allelic single-nucleotide variant present in gnomAD inside
a detected switch is ingested: $\mathbf{157}$ variants, $100\,\%$
non-coding (66\% intron, 14\% intergenic, 13\% regulatory-region,
8\% 5$'$-UTR, $0\%$ missense). Every variant is scored twice,
once by the PWM cross-correlation of Eq.~(\ref{eq:xcorr}) and once
by the AlphaGenome deep-learning model \cite{gankin2026alphagenome}; all
$157$ scoring calls succeeded. The scanner returns $\mathbf{21}$
switches across $\mathbf{17/52}$ genes ($9$ perfect, $6$ gap,
$6$ step). A permutation test finds the loss-of-binding signal
strongly enriched in perfect-type switches
($p=\mathbf{3\!\times\!10^{-4}}$; gap $p=0.55$; step $p=1.0$),
matching the biochemistry of \cite{yamamoto2009differential,ota2013evolutionarily} which
shows that HSF1 prefers continuous nGAAn arrays.

\subsection{Two independent scorers do not agree}
An honest sanity check: across the $157$ variants, $\Delta$PWM and
the AlphaGenome mean score correlate only weakly
(Spearman $\rho=+0.079$, $p=0.33$). This is not a bug. AlphaGenome
was trained on ENCODE and GTEx tracks acquired at $37\,^\circ$C. At
$37\,^\circ$C, HSF1 is a monomer bound to Hsp90 and the switch is
unoccupied, so a variant that abolishes the PWM match cannot
register in any measured track. The temperature-aware Layer~2 of
the cascade is doing work that no purely sequence-to-function model
can deliver on its own. The weak correlation suggests that the PWM-based and AlphaGenome
scores capture largely distinct aspects of variant effects, motivating
the temperature-aware cascade rather than reliance on either sequence-level score alone.

\section{Results}

\begin{figure}[t]
\centering
\includegraphics[width=\linewidth]{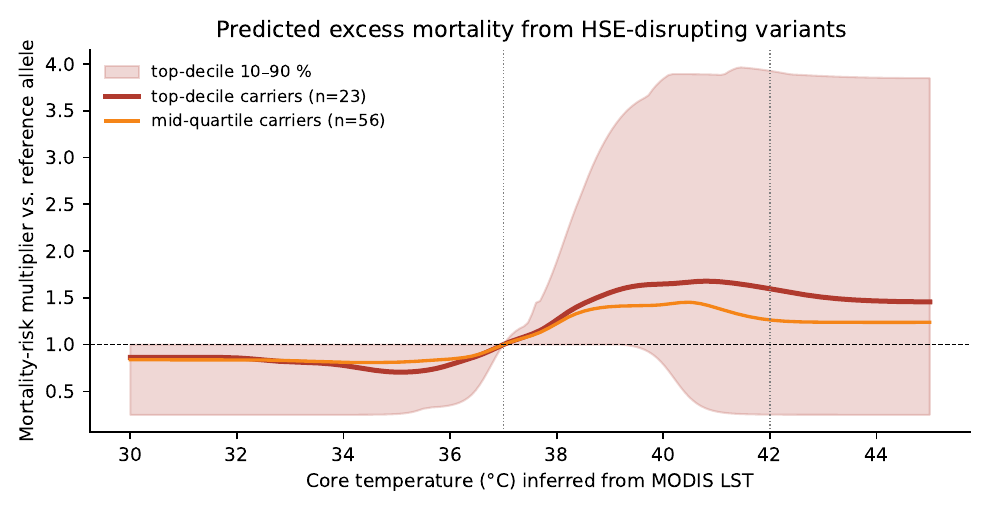}
\caption{Personalised mortality-risk multiplier. Red: mean over
$23$ top-decile switch-disrupting variants. Orange: mean over
$56$ mid-quartile carriers. Pink band: 10 to 90\,\% person-to-person
spread of the top-decile group.}
\label{fig:mortality}
\end{figure}

\begin{table}[t]
\centering
\caption{Heat-response gene panel (52 genes) grouped by functional category.}
\label{tab:panel}
\footnotesize
\scalebox{0.8}{
\begin{tabular}{lp{5.2cm}}
\toprule
Category & Genes \\
\midrule
Cytosolic HSPs / co-chaperones &
HSPA1A, HSPA1B, HSPA1L, HSPA6, HSPA8, HSPA5, HSP90AA1, HSP90AB1,
HSPB1, HSPB8, HSPH1, HSPD1, HSPE1, DNAJA1, DNAJB1, DNAJB4, DNAJB6,
DNAJB9, DNAJC7, BAG3, CRYAB \\
Proteostasis / autophagy &
SQSTM1, NFE2L2, ATG7, ATG12, BECN1 \\
HSF regulators / partners &
HSF1, HSF2, HSF4, HSBP1 \\
Chromatin \& transcription &
JMJD6, JUN, FOS, EGR1, HDAC1, KAT2B \\
Ribosome / translation &
EIF4A1, EIF2S1, RPL13A \\
Climate-physiology (Tier~2) &
G6PD, NR3C2, SCNN1A, SCNN1B, AQP2, AQP3, TRPM4, TRPV1, ATP1A1,
RYR1, CAPN3, HBB, MYH7 \\
\bottomrule
\end{tabular}}
\end{table}
Figure~\ref{fig:mortality} is the paper's central quantitative
output. The horizontal axis is the core body temperature inferred
from MODIS LST by Eq.~(\ref{eq:tcore}); the vertical axis is the
mortality-risk multiplier relative to a person who inherited a
normal switch. A value of $1.0$ means no extra risk.

Three things are visible:
\begin{itemize}
\item Below the normal $37\,^\circ$C baseline the curves lie at or
      below $1.0$. Genetics does not matter when it is not hot.
\item Above $37\,^\circ$C the curves diverge sharply. The switch is
      designed to fully turn on around $40$--$42\,^\circ$C, which is
      exactly where the red curve peaks. A weakened switch fails at
      the moment it is most needed.
\item Peak extra risk is $\mathbf{1.68\times}$ for the strongly
      affected group ($23$ variants) and $1.45\times$ for the
      moderately affected group ($56$ variants) between $40$ and
      $42\,^\circ$C. The 10 to 90\,\% person-to-person spread of the
      top-decile group is $[1.00,\,3.97]\times$.
\end{itemize}

The headline numbers of the whole pipeline are collected for
reference in Table~\ref{tab:results}.

\begin{table}[t]
\centering
\caption{Summary of HSE detection, variant scoring, and predicted mortality-risk results.}
\label{tab:results}
\footnotesize
\begin{tabular}{lr}
\toprule
Quantity & Value \\
\midrule
Genes in panel                        & 52 \\
Genes with $\ge 1$ HSE array                   & 17 \\
HSE arrays detected (P / G / S)                & 9 / 6 / 6 \\
Null hits per $100$\,kb random DNA             & 1 \\
Rare SNVs inside HSE footprints                & 157 \\
Median $|\Delta\mathrm{PWM}|$ (bits)            & 5.73 \\
Max $|\Delta\mathrm{PWM}|$ (bits)               & 11.51 \\
Permutation $p$ ($|\Delta\mathrm{PWM}|$; P / G / S) & $3{\times}10^{-4}$ / $0.55$ / $1.0$ \\
AlphaGenome variants scored                    & $157/157$ \\
Median $|$AlphaGenome score$|$                 & $1.0\!\times\!10^{-4}$ \\
Spearman $\rho$ (PWM vs.\ AlphaGenome mean)     & $+0.079$ ($p=0.33$) \\
Spearman $\rho$ (PWM vs.\ AlphaGenome RNA)      & $+0.168$ ($p=0.035$) \\
Peak mortality multiplier (top-decile)         & $1.68\times$ \\
Peak multiplier 10--90\,\% band                 & $[1.00,\,3.97]\times$ \\
Peak multiplier (mid-quartile)                 & $1.45\times$ \\
\bottomrule
\end{tabular}
\end{table}

\section{Discussion}

The pipeline reads exactly like a sensor-fusion problem. One channel
is an environmental signal at $1\,$km $\times$ daily resolution
from a satellite. The other is a patient-specific calibration (the
switch strength) that is fixed for life and can be measured once
from a swab. The two multiply into a real-time personalised risk
score that a public-health department can, in principle, use to
prioritise cooling-centre outreach, hydration reminders and in-home
checks during a heatwave alert.

\subsection{Where the SP contribution is}
Three of the four operators (Steps 1, 2, 4) are borrowed. The one
that carries the paper is Step 3: writing the DNA scan as a 1-D
cross-correlation with a log-odds kernel makes clear that the
gap- and step-type architectures are exactly the alternating
run-length constraint families a strict-regex scanner drops. Once these constraints are included, the scanner identifies HSE arrays in
17/52 genes, and the permutation test on $|\Delta\mathrm{PWM}|$ shows
significant enrichment for perfect-type HSEs ($p = 3\times10^{-4}$).

\subsection{Limits}
(i) The PWM was trained at normal body temperature; the sequence
score is assumed to transfer to the heat-activated regime because
the switch is a permanent feature of the DNA. (ii) The Arrhenius
energy $E_a$ is a single global value; tissue-specific values
(kidney, skin, cardiac muscle) are known but not yet fitted.
(iii) The satellite temperature is a proxy for what the person
actually feels; indoor cooling, humidity and clothing all modify it.
(iv) Cohort validation of the mortality prediction requires linked
LST-genotype-outcome data, which we do not yet possess. The KORA
F4/FF4 cohort and cross-species mouse HSF1 ChIP-seq
\cite{vihervaara2013transcriptional, mendillo2012hsf1} are the natural next validation targets.

\section{Conclusion}

Heat harms health, but it does not harm everyone equally. Using two
independent signals (one measured by a satellite from $700\,$km
above the person, one measured once from a swab) and four elementary
SP operators, we produce a personalised risk multiplier that peaks
at $1.68\times$ during intense heat. The pipeline fits naturally into
the way a signal-processing engineer already thinks about sensor
fusion.

\section*{Compliance with Ethical Standards}
This work is a purely computational secondary analysis of publicly
available data (Ensembl~112, gnomAD~v4, JASPAR MA0486.2, MODIS
MOD11A1, ERA5-Land, CHIRTS-daily) and does not involve any
prospective research on human or animal subjects performed by any
of the authors. No individual-level identifiable data were used.
Accordingly, no ethical approval was required for this study.
The authors declare no competing interests.

\bibliographystyle{IEEEtran}
\bibliography{references}

\end{document}